\documentclass[twocolumn,pre,twoside,showkeys,floatfix]{revtex4-1}
\usepackage{graphics}
\usepackage{graphicx}
\usepackage{amsmath}
\usepackage{amssymb}
\usepackage{textcomp}
\usepackage{color}
\usepackage{ulem}
\newcommand{\be}{\begin{equation}}
\newcommand{\ee}{\end{equation}}

\newcommand{\dps}{\displaystyle}

\begin{document}

\title{Three-phase contact line and line tension of electrolyte 
solutions\\in contact with charged substrates}

\author{Ingrid Ibagon}
\author{Markus Bier}
\email{bier@is.mpg.de}
\author{S.\ Dietrich}
\affiliation
{
   Max-Planck-Institut f\"ur Intelligente Systeme, 
   Heisenbergstr.\ 3,
   70569 Stuttgart,
   Germany and
   IV. Institut f\"ur Theoretische Physik,
   Universit\"at Stuttgart,
   Pfaffenwaldring 57,
   70569 Stuttgart,
   Germany
}

\date{9 November 2015}

\begin{abstract}
The three-phase contact line formed by the intersection of a 
liquid-vapor interface of an
electrolyte solution with a charged planar substrate is studied in terms 
of classical density
functional theory applied to a lattice model. The influence of 
the substrate  charge
density and of the ionic strength of the solution on the 
intrinsic structure of the three-phase contact line and on the
corresponding line tension is analyzed. We find a negative line tension 
for all
values of the surface charge density and of the ionic strength 
considered. The strength of the line tension 
decreases upon decreasing the contact angle via varying either 
the
temperature or the substrate charge density. 
\end{abstract}

\keywords{line tension, 
          structure of three-phase contact line,  
          electrolyte solutions,
          charged substrates}

\maketitle

\section{Introduction}

The line tension is the free energy per length associated with the contact
line where three phases meet in space. For example, for a sessile liquid
drop on a solid substrate and surrounded by the vapor phase, the contact
line corresponds to the periphery of the circle where the liquid-vapor
interface meets the substrate. 
Although its magnitude is small (both theoretical and experimental values
are of the order of $10^{-12}$ to $10^{-10}$ N \cite{Getta1998, Qu, Dobbs, 
Schneemilch, Pompe, Mugele2002, Takata, Dussaud, Djikaev}), the line 
tension plays an important role for various systems and phenomena such as
spreading of droplets \cite{deGennes, Fan}, wetting of nanoporous surfaces
\cite{Raspal}, stability of emulsions and foams \cite{Aveyard}, drop size
\cite{Weijs}, and many others. 
The line tension has been the subject of numerous theoretical and experimental
investigations (see, e.g., Refs. \cite{SND, IndekeuLT1, IndekeuLT2, 
Rusanov} and references therein). 
Experimental setups to study line tensions encompass solid-liquid-gas 
systems, such as drops on solid substrates \cite{Pompe, Mugele2002}, 
bubbles on solid substrates \cite{Platikanov, Rodrigues} or on particles 
at liquid-gas interfaces \cite{Scheludko, Aveyard1995,Yakubov}, and 
liquid-liquid-vapor
systems, such as liquid lenses at liquid-gas interfaces \cite{Takata,
 Dussaud, Aveyard1999}.
Theoretical investigations include extensions of capillarity theory 
\cite{Rowlinson}, 
which take into account line
tension effects \cite{Widom1995, Marmur}, microscopic theories 
\cite{Tarazona, Getta1998, Bauer,Weijs2011,Zeng2011}, as well as molecular 
dynamics \cite{Bresme1998, Bresme1999, Werder, Hirvi, Schneemilch, Dutta2011,
Ramirez2014} and Monte Carlo \cite{Djikaev, Kulmala, Block} simulations. 
Except for simulations of pure water, these investigations deal with
simple fluids or binary liquid mixtures thereof. 
However, most actual fluids used in wetting applications comprise several
components. 
For polar fluids, such as water, these additional components often carry
an ionic character.
It is well-known that the presence of ions in a fluid creates the Debye
length as an additional length scale, which increases upon decreasing the 
ionic strength.
For dilute electrolyte solutions it typically exceeds the bulk
correlation length of the pure solvent.
Hence the structure and the wetting behavior of an electrolyte solution can
be expected to differ significantly from that of the pure solvent 
\cite{Bier2012,Ibagon,Ibagon2}.

There are only few studies concerned with the influence of electrostatic 
interactions on the line tension \cite{Digilov,Chou,Buehrle,Kang2,Kang,
Dorr2014}.
In Ref.~\cite{Digilov} the theory of capillarity has been extended
taking into account line contributions as well as electric charges at the 
interfaces and at the 
three-phase contact line (TPCL). Within this approach electrowetting has 
been interpreted as a line
tension
effect, but some of the corresponding predictions are in disagreement 
with experimental data \cite{Quilliet}.
In Ref.~\cite{Chou} an equation for the contact angle as function of the 
electrostatic potential at the TPCL and an estimate for the electrostatic
contribution to the line tension have been derived using a variational
approach for a wedge-like geometry. 
Based on a Poisson-Boltzmann theory the analysis in Refs.~\cite{Kang,Kang2}
considers only the electrostatic part of the free energy. 
Therefore, only the electrostatic contribution to the line tension is 
analyzed. 
The density distribution of a conductive liquid close to the three-phase
contact line has been calculated numerically  in Ref.~\cite{Buehrle}, but
the line tension was not studied.  
Recently, D\"orr and Hardt \cite{Dorr} studied the electric double layer
structure close to the TPCL by solving the linearized Poisson-Boltzmann
equation in a wedge geometry, without calculating  the line tension. 
Following the method used in Ref.~\cite{Dorr}, Das and Mitra calculated the
Maxwell stress and the contact angle of drops or bubbles on a charged 
substrate, again without taking into account line effects \cite{Das}.
More recently, D\"orr and Hardt \cite{Dorr2014} computed the line tension
of an electrolyte in contact with a charged substrate by considering a wedge
geometry similar to Refs.~\cite{Kang,Kang2,Dorr,Das}. 
Similarly to Refs.~\cite{Kang,Kang2}, they considered only the electrostatic
contribution to the line tension. 
However, their model differs from the one in Refs.~\cite{Kang,Kang2} in that
it incorporates the deformation of the fluid-fluid interface near the TPCL 
relative to planar shapes.
To our knowledge there are no microscopic calculations of line tensions in 
electrolyte solutions in which both solvent and ion contributions are taken
into account simultaneously.

Here we present a microscopic calculation of the line tension 
and of the intrinsic TPCL structure for a lattice model of an 
electrolyte solution in contact with a charged substrate which takes into 
account solvent and ion contributions via classical density functional theory. 
The wetting phenomena of this model have already been studied in Ref. 
\cite{Ibagon}. In Sec. \ref{MDFT} we recall this model and the 
corresponding 
density functional. The results for the line tension and the TPCL structure for 
both the salt-free solvent and the electrolyte solutions are discussed in Sec. 
\ref{TPCLandLT}. We conclude and summarize our main results in Sec. 
\ref{CS}.

\section{Model and density functional theory}\label{MDFT}

\subsection{Model}\label{model}
We study a semi-infinite lattice model for an electrolyte solution in
contact with a charged wall. This model is the same as the one used in Ref. 
\cite{Ibagon}. It consists
of three components: solvent $(0)$, anions
$(-)$, and cations $(+)$. The $z$ axis is perpendicular to the wall. The 
region
above the wall, which is the one accessible
to the electrolyte
components, 
is divided into a set of cells the centers of which form a simple cubic lattice 
$\{\bf r\}$
with
lattice constant
$a$. The volume $a^3$ of such a cell corresponds roughly to the volumina
of the particles, which are
assumed to be of similar size. The centers of the molecules in the top layer of
the substrate form the
plane $z=0$. At closest approach the centers of the solvent molecules and ions 
are at $z=a$. The
plane
$z=a/2$ is taken to be the surface of the planar wall. 
Each cell is either empty or occupied by a single particle. This mimics the 
steric hard core
repulsion between all particles. Particles at different
sites
interact among each other via
an attractive nearest-neighbor interaction of strength $u$ which is taken to be
the same for all
pairs of particles. In addition, ion pairs interact via the Coulomb potential. 
 
The wall attracts particles only in the first adjacent layer via an interaction 
potential of strength $u_{\text w}$ which is the same for all species. 
In addition it can carry a surface charge density $\tilde\sigma=\sigma ea^{-2}$
which is taken to be localized in the plane $z=a/2$ and which interacts 
electrostatically with the ions; $e>0$ is the elementary charge. 
The surface charge density $\tilde\sigma$ is assumed to be laterally uniform
and independent of the structure of the adjacent fluid, i.e., it is the same for
a liquid-wall and for a gas-wall interface.
This situation is typically realized in EWOD (electrowetting on dielectrics)
setups \cite{Berge1993,Mugele2005}, in which the wall is composed of an 
electrode covered by a micron-sized isolating dielectric layer.
For these systems the areal wall charge density is determined 
by the laterally uniform capacity of the dielectric layer and not by 
charge-regulation mechanisms or by the electric double layer structure in
the electrolyte solution.

For the present model it is known \cite{Ibagon} that in the presence of 
ions a first-order wetting transition occurs to which a prewetting line is 
attached from which layering transition lines depart towards higher 
temperatures (see Fig.~7 in Ref.~\cite{Ibagon}).
These latter artifacts due to the lattice model used here are not expected to
influence the results below because in the following only bulk states at
liquid-vapor coexistence below the wetting transition are considered.

\subsection{Density functional}\label{dft}
We denote the dimensionless lattice positions as ${\bf \bar r}={\bf r}/a$ and 
$\tilde \rho_i({\bf
\bar r})=\rho_i({\bf \bar r})a^{-3}$ with $i\in\{0,+,-\}$ denotes the
number densities of the solvent ($i=0$) and of the $\pm$-ions. The 
equilibrium profiles $\rho_0$, 
$\rho_+$, and $\rho_-$ minimize the following grand canonical density 
functional:

\be\label{functional1}
\begin{split}
 \beta\Omega\left[\{\rho_i({\bf \bar r})\}\right]&=\sum_{{\bf \bar r}}\left[
\sum_i\rho_i({\bf \bar r})\ln{\rho_i({\bf \bar r})}\right.
\\&\left.+\Big(1-\sum_i\rho_i({\bf \bar
r})\Big)\ln{\Big(1-\sum_j\rho_j({\bf \bar r})\Big)}\right]\\
&+\frac12\beta\sum_{\substack{{\bf \bar r},{\bf \bar r'}\\ {\bf \bar r}\neq {\bf
\bar r'}}}\sum_{i,j}\rho_i({\bf \bar r})\rho_j({\bf \bar r'})w\left(|{\bf \bar 
r} -{\bf
\bar r'}|\right)
\\&-\beta\sum_{
{\bf \bar r}}\sum_iu_{\text
w}\delta_{\bar z, 1}\rho_i({\bf \bar r})-\beta\sum_{{\bf \bar 
r}}\sum_i\mu_i\rho_i({\bf \bar
r})
\\&+2\pi l_{B}\int_V\!d^3\bar
r^*\frac{\left({\bf
D}\left({\bf \bar r^*},[\rho^*_{\pm}]\right)\right)^2}{\varepsilon(\rho^*_0({\bf
\bar
r^*}))},
\end{split}
\ee
where $\beta=(k_BT)^{-1}$ is the inverse thermal energy; $\mu_i$ is the 
chemical potential
of species $i$; $\tilde l_{B}=l_Ba=e^2\beta/(4\pi\varepsilon_0)$ is the
Bjerrum length in vacuum; and ${\bf \bar r^*}={\bf r^*}/a$,  
$\rho^*_i({\bf \bar r^*})=\rho_i({\bf \bar r})$ for all ${\bf \bar r^*}\in 
\mathbb{R}^3$ and ${\bf \bar r}\in\mathbb{Z}^3$ with
$\max\left(|\bar x^*-\bar x|, |\bar y^*-\bar y|, |\bar z^*-\bar z|\right)\le 
1/2$, i.e., with ${\bf\bar r}$ corresponding to that site of the 
discrete cubic lattice $\mathbb{Z}^3$ being located closest to position
${\bf\bar r^*}$ in the continuous space $\mathbb{R}^3$. 
The pair potential common for all species is $w\left(|{\bf r}-{\bf r'}|\right)
=-u$ for nearest neighbors (i.e., $u>0$ corresponds to 
attraction) and
$w\left(|{\bf r}-{\bf r'}|\right)=0$ beyond;  $-u_{\text w}$ is the strength of 
the
attractive ($u_{\text w}>0$) substrate potential acting on the first
layer $z=a$. $\tilde
{\bf D}={\bf D}ea^{-2}$ is the actual electric
displacement generated by the ions and
the surface charge
density $\tilde\sigma=\sigma ea^{-2}$, satisfying Gau\ss's law \cite{Jackson} 
with the dimensionless gradient $\nabla$ obtained by rescaling with $a$:
\be
\nabla \cdot {\bf D}\left({\bf\bar
r^*},[\rho^*_{\pm}]\right)=\sum_iq_{i}\rho^*_i({\bf \bar r^*})+\sigma\delta(\bar
z-1/2)\label{GL3D};
\ee
The concept underlying this form of Gau\ss's law is that all microscopic
charges besides the ionic monopoles and the surface charges, e.g., those due to
permanent or induced dipoles, are implicitly accounted for in terms of the relative
permittivity $\varepsilon(\rho^*_0({\bf \bar r^*}))$. 
Here the relative permittivity is assumed to be dominated by the solvent properties,
as it is the case for polar solvents such as water, so that it depends only on the 
solvent density $\rho_0({\bf\bar r})$ but not on the ion densities 
$\rho_\pm({\bf \bar r})$.

The description in Eq.~(\ref{functional1}) does not account for the 
structure of a hydration shell, neither in the bulk nor at interfaces.
For actual systems there is a strong dependence of, e.g., the value of the 
interfacial tension of a liquid-vapor interface on atomistic details of ion
hydration \cite{Tobias2013}; but it is not the aim of the present study to
model such details.
Moreover, here the solubility of ions is accounted for merely effectively via
the ion-solvent interaction $w\left(|{\bf \bar r} -{\bf \bar r'}|\right)$, 
which, for the sake of simplicity, is the same between all particle species.
More realistic descriptions could be used instead, e.g., in terms of the Born
energy \cite{Born1920}, but this is not done here for reasons of simplicity.

The bulk phase diagram, i.e., the solvent and the $\pm$-ion densities in
the liquid ($\{\rho_{i,l}\}$) and in the gas phase 
($\{\rho_{i,g}\}$) of the solution, has already been 
determined in Ref.
\cite{Ibagon}. The bulk equilibrium densities are calculated by minimizing the 
bulk grand canonical
potential
\begin{multline}\label{bulk1}
\frac{\beta \Omega[\{\rho_i\}]}{\bar V}=\rho_0(\ln{\rho_0}-\mu^*_0)
+I(2\ln{I}-\mu^*_I)\\
+(1-\rho_0-2I)\ln{(1-\rho_0-2I)}-\frac{1}{T^*}(\rho_0+2I)^2,
\end{multline}
where $I:=\rho_+=\rho_-$ (due to local charge neutrality) is the 
so-called ionic strength for
monovalent ions; $\mu^*_0=\beta\mu_0$,
$\mu^*_I=\beta(\mu_++\mu_-)$, $T^*=\frac{1}{3\beta u}$ is the reduced 
temperature, and $V=\bar Va^3$ is the volume of the fluid. The last
term in Eq.~(\ref{functional1}) vanishes because in the bulk $\mathbf{D}=0$ due 
to Eq.~(\ref{GL3D}).
For $I=0$, the reduced critical temperature is $T^*_c(I=0)=0.5$ and the 
critical number density of the 
solvent is
$\rho_{0,c}(I=0)=0.5$. For $I\neq0$, the reduced critical
temperature $T^*_c$ is independent of $I$ whereas $\rho_{0,c}(I)=0.5-2I$ 
\cite{Ibagon}.

At two-phase coexistence for temperatures below the critical point, 
$T^*\leq T^*_c$, the bulk densities $\{\rho_{i,l}\}$ in the liquid and 
$\{\rho_{i,g}\}$ in the gas phase are fully specified by the \textit{four} values 
$\rho_{0,l}$, $I_l$, $\rho_{0,g}$, and $I_g$, which have to fulfill the \textit{three}
coexistence conditions (see Eq.~(26) of Ref.~\cite{Ibagon})
\begin{align}
   \mu_0[\{\rho_{i,l}\},T^*] &= \mu_0[\{\rho_{i,g}\},T^*], 
   \notag\\
   \mu_I[\{\rho_{i,l}\},T^*] &= \mu_I[\{\rho_{i,g}\},T^*],
   \notag\\
   p[\{\rho_{i,l}\},T^*] &= p[\{\rho_{i,g}\},T^*],
   \label{eq:coexcond}
\end{align}
where $p=-\Omega/V$ is the pressure.
Hence, in addition to Eq.~(\ref{eq:coexcond}), one of the four values 
$\rho_{0,l},I_l,\rho_{0,g},I_g$ can be fixed arbitrarily; in the following
we choose the ionic strength in the liquid, $I_l$, which, is simply called
\textit{the} ionic strength $I$ in Sec.~\ref{TPCLandLT}.

With the bulk properties fixed and with a given expression for the 
relative permittivity
$\varepsilon(\rho^*_0({\bf \bar r^*}))$, the functional in 
Eq.~(\ref{functional1}) can be used to
study the wetting behavior of electrolyte solutions as it has been done
in Ref. \cite{Ibagon}. If the
substrate potential depends only on the direction $\bar z$ perpendicular to the 
wall and if the
boundary
condition for $\bar z\to \infty$ is laterally homogeneous, the 
equilibrium density profiles depend on
$\bar z$ only but not on $\bar x$ or $\bar y$.
Accordingly the grand canonical functional in Eq.~(\ref{functional1}) 
decomposes into the bulk
contribution given by Eq.~(\ref{bulk1}) and into a surface contribution 
proportional to the area
$A=\bar A a^2$ of the substrate:
\be
\Omega\left[\{\rho_i({\bf \bar 
r})\}\right]=V\Omega_b[\rho_i]+A\Omega_s[\{\rho_i(\bar z)\}].
\ee
The substrate-liquid surface tension $\gamma_{s,l}$ and the substrate-gas 
surface tension
$\gamma_{s,g}$ are given by
\be\label{slsurfacetension}
\gamma_{s,l}=\min_{\{\rho_i(\bar z)\}}{\Omega_s[\{\rho_i(\bar z)\}]},\ \ \ \
\text{for}\ 
\rho_i(z\to\infty)=\rho_{i,l},
\ee
and
\be\label{sgsurfacetension}
\gamma_{s,g}=\min_{\{\rho_i(\bar z)\}}{\Omega_s[\{\rho_i(\bar z)\}]},\ \ \ \ 
\text{for}\ 
\rho_i(z\to\infty)=\rho_{i,g},
\ee
respectively. The equilibrium density profiles which minimize 
Eq.~(\ref{slsurfacetension})
correspond to the substrate-liquid interface, while 
the ones which minimize Eq.~(\ref{sgsurfacetension}) correspond
to the substrate-gas
interface, which consists of the emerging substrate-liquid and liquid-gas 
interfaces
separated by a
liquid-like layer of thickness $\tilde\ell_0(T)=a\ell_0(T)$. At two-phase 
coexistence, i.e., for
$\mu^*_{i}=\mu^*_{i,co}$, both density profiles described above are 
equilibrium density
distributions.

\begin{figure}[!t]
\begin{center}
\includegraphics[scale=1.1,clip=true]{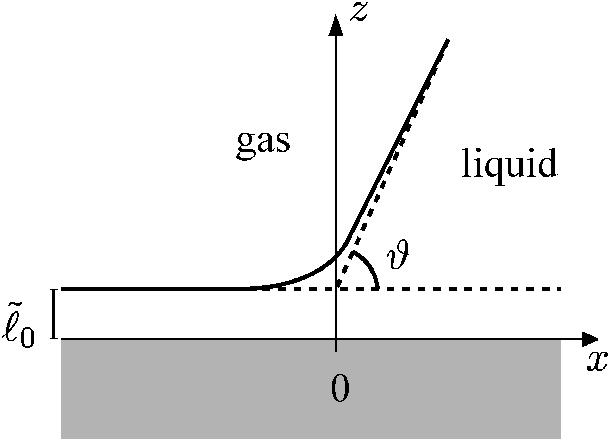}
\end{center}
   \caption{Generic shape $z=\tilde\ell(x)$ of the liquid-gas interface  
            (full line) near the three-phase contact line (TPCL) on a
            homogeneous planar surface; $\tilde\ell_0$ is the equilibrium 
            thickness of the microscopicly thin liquid-like film for $T$ 
            below the wetting transition temperature $T_{\text w}$ and 
            at two-phase coexistence. 
            The dashed lines represent the asymptotes of $z=\tilde\ell(x)$ 
            for $x\to\infty$ and $x\to-\infty$. 
            The position $x=0$ is defined as the point at which the 
            asymptotes intersect. 
            The asymptote $z=\tilde\ell(x\to\infty)$  intersects the 
            substrate with the contact angle $\vartheta$. 
            The density distribution is translationally invariant along
            the $y$ direction perpendicular to the $x$-$z$-plane.}
\label{lxsketch}
\end{figure}

Imposing these two distinct boundary
conditions at $\bar x=\pm\infty$ for
$\bar z\to\infty$, i.e., $\rho_i(\bar x\to\infty,\bar z\to\infty)=\rho_{i,l}$ 
and
$\rho_i(\bar x\to-\infty,\bar z\to\infty)=\rho_{i,g}$, the minimization of 
Eq.~(\ref{functional1})
leads to an equilibrium density
distribution $\rho_i(\bar x, \bar z)$ which interpolates smoothly between a 
substrate-gas interface
at
$\bar x\to-\infty$ and a substrate-liquid interface at $\bar x\to+\infty$. A 
specific definition of the
local position of the liquid-gas interface renders a curve $\bar z=\ell(\bar 
x)$ (see, c.f., Eq.~(\ref{lx}))
such that $\ell(\bar x\to-\infty)=\ell_0(T)$ and 
$\ell(\bar x\to\infty)=\ell_0(T)+\bar x\tan\vartheta$, where $\vartheta$ is the
contact
angle (see Fig.
\ref{lxsketch}). This arrangement leads to the formation of a straight 
TPCL independent of $\bar y$ where
the liquid-gas, the substrate-gas, and the substrate-liquid interfaces
intersect.

For  $\rho_i({\bf \bar r})=\rho_i(\bar x,\bar z)$, the density functional in 
Eq.~(\ref{functional1}) can be written as
\begin{widetext}
\be\label{functional2D}
\begin{split}
 \frac{\beta\Omega\left[\{\rho_i(\bar x,\bar z)\}\right]}{\bar 
L}&=\mkern-18mu\sum_{\bar
x=-n_x}^{n_x}\sum_{\bar z=1}^{\bar L_z}\left\{
\sum_i\rho_i(\bar z)\ln{\rho_i(\bar x,\bar z)}
+\Big(1-\sum_i\rho_i(\bar x,\bar z)\Big)\ln{\Big(1-\sum_j\rho_j(\bar x,\bar
z)\Big)}\right. \\
&\phantom{=}-\frac{\beta u}{2}\sum_{ij}\rho_{i}(\bar x,\bar 
z)\left(\rho_{j}(\bar x\!+\!1,\bar
z)+\rho_{j}(\bar x\!-\!1,\bar
z)+\rho_{j}(\bar x,\bar z\!+\!1)\right.\\
&\phantom{=}\left.\left.+\rho_{j}(\bar x,\bar 
z\!-\!1)+2\rho_j(\bar x, 
\bar z)\right)-\beta u_{\text w}\sum_i\rho_i(\bar x,\bar z)\delta_{\bar 
z,1}
-\beta\sum_i\mu_i\rho_i(\bar x,\bar z)\right\}\\
&\phantom{=}+2\pi
l_{B}\int_{-\bar L_x/2}^{\bar L_x/2}\int_{1/2}^{\bar L_z+1/2}\!d\bar x^*d\bar
z^*\frac{\left(\mathbf{D}(\bar x^*,\bar
z^*,[\rho^*_{\pm}])\right)^2}{\varepsilon(\rho^*_0(\bar x,\bar z))},
\end{split}
\ee
\end{widetext}
where $n_x=\frac{\bar L_x-1}{2}$ with lateral system size $\bar L_x$, 
$\bar L=L/a$ is the contact line length in the invariant $\bar y$ 
direction, and the fluid volume is
$V=L_xL_zL$; $i,j=0,+,-$. 

Gau\ss's law (Eq.~\ref{GL3D}) can be written as
\be
\nabla \cdot {\bf D}\left({\bar x^*, \bar 
z^*},[\rho^*_{\pm}]\right)=\sum_iq_{i}\rho^*_i(\bar
x^*, \bar z^*).
\ee
with the boundary conditions 
\be\label{BC2D}
\begin{aligned}
\left.D_z(\bar x^*, \bar z^*, [\rho^*_\pm])\right|_{\bar z^*=1/2}&=\sigma,\\
\left.D_z(\bar x^*, \bar z^*, [\rho^*_\pm])\right|_{\bar z^*=\bar L_z+1/2}&=0,\\
\left.D_x(\bar x^*, \bar z^*, [\rho^*_\pm])\right|_{\bar x^*=-\bar L_x/2}&=0,\\
\left.D_x(\bar x^*, \bar z^*, [\rho^*_\pm])\right|_{\bar x^*=\bar L_x/2}&=0,
\end{aligned}
\ee
which follow from the overall charge neutrality.

The relative permittivity $\varepsilon(\bar z^*)$ is taken to depend locally on 
the solvent density
$\rho^*_0(\bar x^*,\bar z^*)$ through the Clausius-Mossotti expression 
\cite{Jackson}
\be\label{epsi2D}
\varepsilon(\rho^*_0(\bar x^*,\bar 
z^*))=\frac{1+\frac{2\alpha}{3\varepsilon_0}\rho^*_0(\bar x^*,\bar
z^*)}{1-\frac{
\alpha}{
3\varepsilon_0 }
\rho^*_0(\bar x^*,\bar z^*)},
\ee
where $\alpha$ is an effective polarizability of the solvent molecules. In the 
following its
value is chosen such that $\varepsilon=60$ for $\rho_0=1$; this choice 
corresponds to a
mean value for liquid water along the liquid-gas coexistence curve.
It is well-known that the Clausius-Mossotti relation between the relative
permittivity $\varepsilon$ and the polarizability $\alpha$ holds only for
dilute gases.
However, Eq.~(\ref{epsi2D}) is merely used as a simple functional form in 
order to obtain the dependence on $\rho^*_0$ of the relative permittivity 
with $\alpha$ being a fitting parameter which is adapted to interpolate 
between $\varepsilon(\rho^*_0)\to1$ for $\rho^*_0\to0$ and a large value
(here $\varepsilon(\rho^*_0)\to60$) for $\rho^*_0\to1$.

The Euler-Lagrange equations, which follow from the minimization of Eq.
(\ref{functional2D}) analogously to the procedure presented in Subsec. IIC in 
Ref.~\cite{Ibagon}, are given by 
\begin{widetext}
\begin{multline}\label{ELE2D}
  \ln{\rho_i(\bar x,\bar z)}-\mu^*_i-\beta u_{\text w}\delta_{\bar z,1}
-\ln{\Big(1-\sum_j\rho_j(\bar x, \bar
z)\Big)}\\
-\frac{1}{3T^*}\sum_j\left(2\rho_{j}(\bar x,\bar z)+\rho_j(\bar x+1,\bar 
z)+\rho_j(\bar x-1,\bar
z)+\rho_{j}(\bar x,\bar z+1)+\rho_{j}(\bar x,\bar z-1)\right)\\
+q_i\int_{\bar z-1/2}^{\bar z+1/2}\!d\bar z^*\phi(\bar x^*,\bar z^*)-2\pi
l_{B}\int_{\bar z-1/2}^{\bar z+1/2}\!d\bar z^*\frac{\left(\mathbf{D}(\bar 
x^*,\bar
z^*,[\rho^*_{\pm}])\right)^2}{\left(\varepsilon(\rho^*_0(\bar x^*,\bar
z^*))\right)^2}\varepsilon'\left(\rho^*_0(\bar x^*,\bar
z^*)\right)\delta_{i,0}=0,
\end{multline}
\end{widetext}
with $i,j=0,+,-$, where $q_ie$ is the electric charge of component $i$,
$T^*=\frac{1}{3\beta u}$ is the reduced temperature and $\mu^*_i=\beta \mu_i$. 
$\phi(\bar
x^*,\bar z^*)=\beta e \tilde \phi(x^*,z^*)$ is the dimensionless
electrostatic potential which fulfills
\be\label{EDEP}
 \mathbf{D}(\bar x^*,\bar z^*)=-\frac{\varepsilon}{4\pi l_{B}}\nabla\phi(\bar 
x^*,\bar z^*).
\ee
At the wall the
convention $\rho_j(\bar x,0)=0$ is
used. 

For given chemical potentials $\mu_{i,co}$ at coexistence, the coupled
equations in Eq.~(\ref{ELE2D})  are solved for $\{\rho_i(\bar x,\bar 
z)\}$ numerically by applying a Picard iteration scheme.
The electrostatic potential $\phi(\bar x^*,\bar z^*)$ is calculated for 
each iteration step by solving Poisson's equation 
\be\label{PE2D}
\nabla\cdot\left(\varepsilon(\rho^*_0(\bar x^*,\bar
z^*))\nabla\phi(\bar
x^*,\bar z^*)\right)=-4\pi l_B\sum_iq_i\rho_i^*(\bar x^*,\bar z^*),
\ee
which is a nonlinear integro-differential equation for $\phi$ 
after eliminating $\rho_\pm^*(\bar x^*,\bar z^*)$ by means of 
Eq.~(\ref{ELE2D}).

\subsection{Line tension calculation}\label{LTC}

The line tension $\tau$ is calculated from the equilibrium 
density profiles 
$\{\rho_i(\bar x,\bar z)\}$
using the following
definition for $\tau$: 
\be\label{lteq}
\Omega=\sum_{\alpha=g,l}V_\alpha\Omega_\alpha+A_{s,g}\gamma_{s,g}+A_{s,l}\gamma_
{ s , l } +A_ { l
,g}\gamma_{l,g}
+\tau
L+\cdots,
\ee
where $V_\alpha$ is the volume of phase $\alpha$ with $\alpha\in\{g,l\}$ and 
$\Omega_\alpha$
is the bulk free
energy
density of this phase; $\gamma_{s,g}$, $\gamma_{s,l}$, and $\gamma_{l,g}$ 
are 
the
interfacial tensions and $A_{s,g}$, $A_{s,l}$, and $A_{l,g}$ the corresponding 
interfacial areas of
the substrate-gas, substrate-liquid, and liquid-gas interfaces, 
respectively. 
$L$ is the length of
the three-phase
contact line, $\tau$ is the line tension and $\cdots$ denotes subleading terms 
which vanish for macroscopically long contact lines
$L\to\infty$.

The
plane $\bar z=0$ is chosen as
the
substrate-fluid dividing interface. In Ref. \cite{SND} it has been proposed 
that in order to
determine
the line tension $\tau$ unambiguously from microscopic calculations in a finite 
box, its boundaries
have to be chosen such that the interfaces are cut perpendicularly and that its 
edges
are placed inside the homogeneous regions of the system. Here, in order 
to 
calculate the line
tension, the
integration box proposed in Ref. \cite{SND} has been used (see Fig.~7 in 
Ref. 
\cite{SND} and, c.f.,
Fig.~\ref{Figctau}). However, in a
lattice model this type of box introduces technical difficulties for the 
integration procedure which
lead to numerical errors (see Appendix \ref{ctau} for more details). Therefore, 
in order to verify
the consistency of the results, the line tension has been calculated for 
various sizes of the
integration box as described in Appendix \ref{ctau}.

\subsection{Choice of parameters}

The values of the parameters used here are the same as the ones
used in Ref.~\cite{Ibagon}. 
The lattice constant $a$ is chosen to be equal to $4\,\text\AA$,
so that the maximal density $1/a^3$ lies between the number densities
for liquid water at the triple point and at the critical point. 
Accordingly, the choice $l_{B}=100$ corresponds to
$T\approx417\,\text{K}$. 
This temperature lies between the triple point temperature of 
$273\,\text{K}$ and the critical point temperature of $647\,\text{K}$
for water. 
In our units $1\,\text{mM}=10^{-3}\,\text{mol}/\text{liter}$ corresponds
to $\rho_i=\tilde\rho_ia^3=3.9\times10^{-5}$. 
The values for the reduced surface charge density $\sigma$ are in the 
range between 0 and $10^{-2}$. 
For $a=4\,\text{\AA}$ the latter value corresponds to 
$1\,\mu\text{C}/\text{cm}^2$, which can be achieved for an EWOD setup
(see Sec.~\ref{model}) by applying the moderate voltage of $60\,\text{V}$ 
across a $100\,\text{nm}$ thick isolating dielectric layer with a typical 
dielectric constant $\epsilon_\text{layer}=2$.
In these units $\beta\tau a=0.1$ corresponds to 
$\tau\approx1.4\times10^{-12}$~N.

\section{Structure of the three-phase contact line and 
line tension}\label{TPCLandLT}

\subsection{Line tension of the pure solvent}

\begin{figure}[!t]
\begin{center}
\includegraphics[width=8cm]{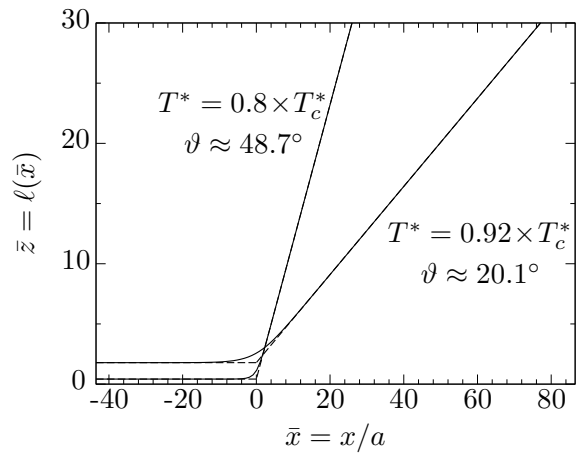}
\end{center}
   \caption{Local positions of the liquid-gas interface for two temperatures
            with $u_{\text{w}}/u=0.69$ for the pure solvent. 
            The system undergoes a second-order wetting transition at 
            $T^*_{\text{w}}\simeq0.95\times T^*_c$ (see Fig.~2(b) in 
            Ref.~\cite{Ibagon}). 
            The positions of the liquid-gas interfaces $\bar z=\ell(\bar x)$
            (full lines) have been calculated from the density profiles 
            $\rho_0(\bar x,\bar z)$ according to Eq.~(\ref{lx}). 
            For both $\bar x\to\infty$ and $\bar x\to-\infty$, the curve
            $\bar z=\ell(\bar x)$ approaches the asymptotes (dashed lines) from
            above. 
            The position $\bar x=0$ is defined as the point at which the 
            asymptotes intersect.}
\label{CPS}
\end{figure}

First, we consider the case $I=0$. As explained in Refs.~\cite{Pandit1982, 
Pandit1983,Ibagon}, in this case  the ratio 
$u_\text{w}/u=3T^*\beta u_{\text{w}}$ controls the wetting and drying 
transitions. 
For $u_{\text w}/u>1$ the substrate is so strong that it is already wet at
$T^*=0$; within the range $0.5<u_{\text w}/u<1$ there is a wetting 
transition at $T^*_{\text w}>0$; and {within} the parameter
range $0\leq u_{\text w}/u<0.5$ a drying transition occurs. 
Here, the liquid-gas interfaces near the TPCL and the line tension  are 
studied for the specific choice $u_{\text{w}}/u=0.69$, for which the system
undergoes a second-order wetting transition (see Fig.~2(b) in 
Ref.~\cite{Ibagon}) at $T^*_{\text{w}}\simeq0.95T^*_c$. 
We note that second-order wetting transitions in a pure solvent with 
short-ranged interactions is not very realistic as most wetting transitions
either are of first order (due to weak van der Waals interactions) 
or comprise a  first-order thin-thick transition followed by a second-order wetting 
transition (due to strong van der Waals interactions) \cite{Bonn2009}.
However, the order of wetting transitions of the pure solvent is not important
here; instead we intend to exploit the technical advantages offered by
short-ranged interactions for the present study.

Figure~\ref{CPS} shows the temperature dependence of the shape $\ell(\bar x)$
of the local liquid-gas interface position defined as 
\be\label{lx}
\ell(\bar x)=\frac{\dps\sum_{\bar z=1}^{\bar L_z}\left(\rho_0(\bar x,\bar
z)-\rho_{0,g}\right)}{\rho_{0,l}-\rho_{0,g}}.
\ee
In the case of second-order wetting transitions, the curve $\bar z=\ell(\bar x)$ 
approaches the
asymptotes for
$\bar x\to\infty$ and $\bar x\to-\infty$ from above (Fig.~\ref{CPS}). This 
result is in
qualitative
agreement with previous ones also obtained in the presence of
second-order wetting transitions \cite{IndekeuLT1,
Getta1998, Bauer, Merath}. 

\begin{figure}[!t]
\begin{center}
\includegraphics[width=8cm]{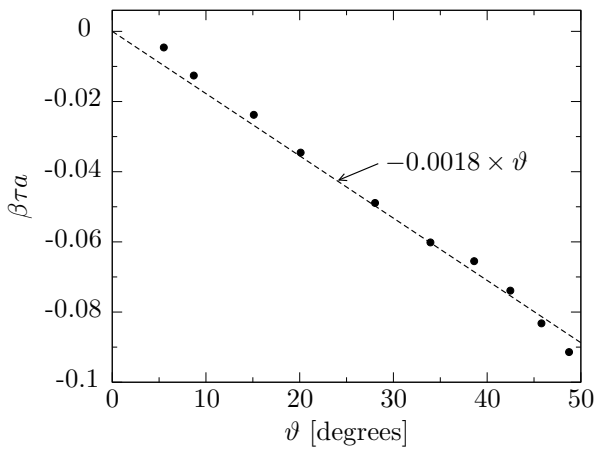}
\end{center}
   \caption{Dependence of the line tension $\tau$ on the contact
            angle for the same pure system as in Fig.~\ref{CPS}. 
            The numerical results for the line tension ($\bullet$) are 
            consistent with the predictions of Ref.~\cite{IndekeuLT1} for
            systems exhibiting second-order wetting transitions with 
            short-ranged interactions, i.e., $\tau$ is negative and for
            $\vartheta\to0$ it vanishes as $\tau\sim-\vartheta$. 
            The uncertainty of the line tension values corresponds 
            approximately to the size of the symbols. The dashed line 
            is a fit.
            For details concerning the calculation of the line tension see
            Appendix~\ref{ctau}.}
\label{ltps}
\end{figure}

The line tension as a function of the contact
angle $\vartheta$ is presented in Fig.~\ref{ltps}. The contact angle has been 
changed by 
varying the
temperature $T^*$. The results
for the line tension are compatible with the prediction of the interface 
displacement model (IDM)
\cite{IndekeuLT1} for a system with short-ranged interactions approaching a 
second-order wetting
transition at two-phase coexistence. In this case, the line tension $\tau$ is 
negative and vanishes
as $\tau\sim-\vartheta$. The order of magnitude of $\beta\tau 
a\approx0.1$, 
which corresponds to
$\tau\approx1.4\times10^{-12}$ N, is comparable also
with values obtained from other theoretical approaches for 
one-component, charge-free fluids \cite{Getta1998, Qu, Dobbs} and from 
computer simulations \cite{Djikaev, Schneemilch} as well as with experimental
results \cite{Dussaud, Pompe, Mugele2002, Takata}.

\subsection{Line tension of an electrolyte solution}

\begin{figure}[!t]
\begin{center}
\includegraphics[width=8cm]{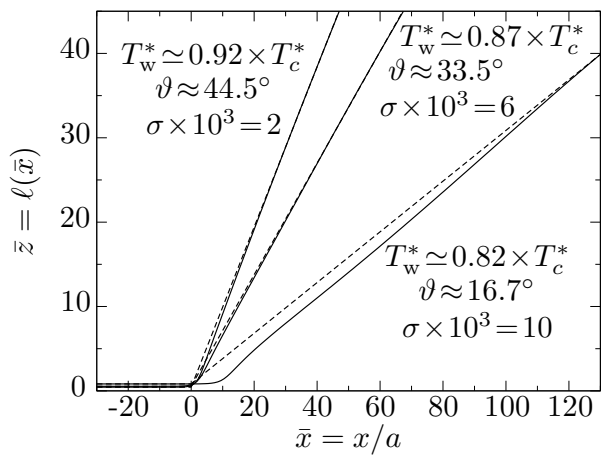}
\end{center}
   \caption{Shapes of liquid-gas interfaces for an electrolyte solution, 
            which exhibits a first-order wetting transition, for various 
            surface charge densities $\tilde\sigma=\sigma ea^{-2}$, fixed
            temperature $T^*=0.8\times T^*_c$, $u_{\text{w}}/u=0.69$, and 
            fixed $I=3.9\times10^{-5}$ ($\tilde I=1\,$mM). 
            Note that for the electrolyte solution the wetting temperature 
            $T_\text{w}$ depends on the surface charge density (see Fig.~5 in
            Ref.~\cite{Ibagon}). 
            The shapes $\ell(\bar x)$ (full lines) have been obtained from
            the density profiles $\rho_0(\bar x,\bar z)$ using Eq.~(\ref{lx}).
            For first-order wetting, the local interface position 
            $\bar z=\ell(\bar x)$ approach their asymptotes (dashed lines) 
            from above for $\bar x\to-\infty$ and from below for 
            $\bar x\to\infty$. 
            The position $\bar x=0$ is defined as the point at which the pair
            of asymptotes intersect.}
\label{ES}
\end{figure}

In this section we study the influence of the ionic strength $\tilde 
I=Ia^{-3}$ and of the surface charge density $\tilde \sigma=\sigma e a^{-2}$
on the TPCL and the line tension. 
As discussed in Refs.~\cite{Ibagon,Ibagon2}, within the chosen lattice
model for an electrolyte solution, if $\sigma\neq0$ and $I\neq0$ the system 
undergoes a first-order wetting transition, irrespective of the order of
the wetting transition of the pure solvent. 
In this case, the wetting transition temperature $T^*_\text{w}$ decreases with
increasing surface charge density $\sigma$ of the substrate for fixed ionic 
strength $I$ or with decreasing ionic strength $I$ for fixed surface charge
density $\sigma$.
Therefore, there are three different routes to vary the contact angle: (i) 
changing the reduced temperature $T^*$ and keeping the surface charge 
density $\sigma$ and the ionic strength $I$ fixed; (ii) changing the surface 
charge density of the substrate $\sigma$ and keeping the temperature $T^*$ and
the ionic strength $I$ fixed; and (iii) changing the ionic strength $I$ and
keeping the temperature $T^*$ and the surface charge density $\sigma$ fixed.
Here we consider the routes (i) and (ii) for two values of the ionic strength: 
$I=3.9\times10^{-5}$ ($\tilde I=1$mM) and $I=3.9\times10^{-4}$ ($\tilde 
I=10$mM) with $u_{\text{w}}/u=0.69$.

\begin{figure}[!t]
\begin{center}
\includegraphics[scale=0.8]{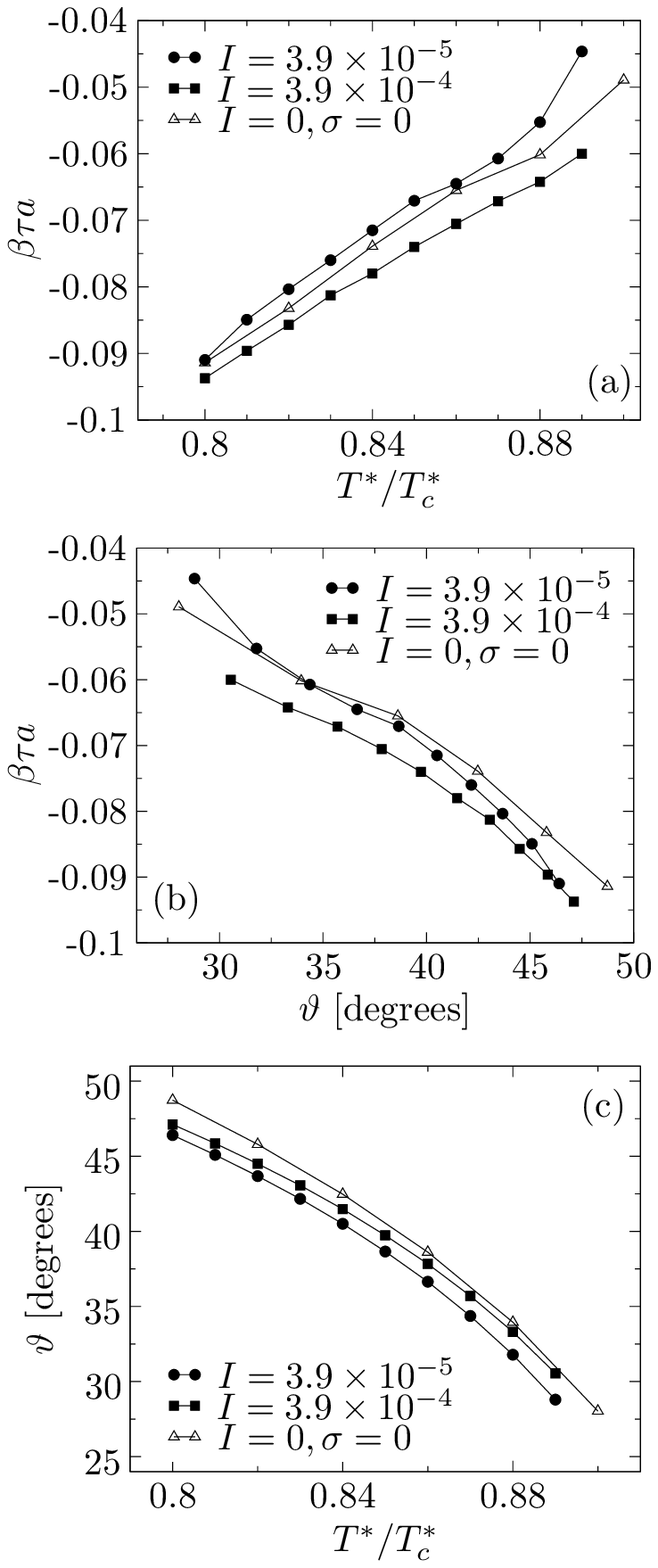}
\end{center}
\caption{Line tension $\tau$ as a function of temperature $T^*$ (a) and of the
         contact angle $\vartheta$ (b) for $\sigma=1\times10^{-3}$ and 
         $u_{\text{w}}/u=0.69$. 
         The two types of full symbols correspond to distinct values of the
         ionic strength $I=\tilde Ia^3$ in the bulk liquid phase ($\bullet$
         for $I=3.9\times10^{-5}$ ($\tilde I=1$mM) and 
         $\scriptstyle\blacksquare$ for $I=3.9\times10^{-4}$ 
         ($\tilde I=10$mM)); the uncertainty
         of the line tension values corresponds approximately to the size of
         the symbols.
         For constant $\sigma$ and $T^*$, the strength $|\tau|$ of the 
         (negative) line tension $\tau$ increases upon increasing the ionic
         strength $I$ (see $\bullet$ and $\scriptstyle\blacksquare$ in panel 
         (a)).         
         The open triangles $\vartriangle$ correspond to the case $I=0$ 
         and $\sigma=0$ (see Fig.~\ref{ltps}), i.e., they differ from
         the filled symbols $\bullet$ and $\scriptstyle\blacksquare$ not only
         with respect to the ionic strength $I$ but also with
         respect to the surface charge density $\sigma$. 
         See Appendix~\ref{ctau} for details concerning the calculation of the
         line tension.
         Panel (b) shows the general trend of an increasing strength
         of the line tension, upon increasing the contact angle 
         $\vartheta$.
         Panel (c) displays the contact angle $\vartheta$ as function of
         $T^*$ with $I$ and $\sigma$ fixed (route (i)).}
\label{tauvsT}
\end{figure}

Figure \ref{ES} shows the shape of the liquid-gas interface as 
obtained from Eq.~(\ref{lx}) 
for fixed
temperature $T^*=0.8T^*_c$, fixed ionic strength $I=3.9\times10^{-5}$ 
($\tilde 
I=1$mM), and for three
different
values of the surface charge density $\sigma$ (route (ii)). If the 
wetting transition is 
first order, the local interface profile
$\bar z=\ell(\bar x)$
 approaches its asymptote from below for $\bar x\to\infty$ and from above for
$\bar x\to-\infty$. For large contact angles, i.e., for small values of 
$\sigma$ (which is in line with the corresponding statement at the 
beginning of the previous paragraph), in Fig.~\ref{ES}, $\bar z=\ell(\bar
x)$ follows its asymptotes closely. The deviation from the 
asymptotes
increases for decreasing contact angles. The behavior of the shape of the
liquid-gas interface is similar for the case
in which the contact angle is changed using route (i). These results for the 
shape of the interface are
in line with those of Refs. \cite{IndekeuLT1, Getta1998, Bauer, Merath} 
for first-order wetting in charge-free fluids.

\begin{figure}[!t]
\begin{center}
\includegraphics[scale=0.8]{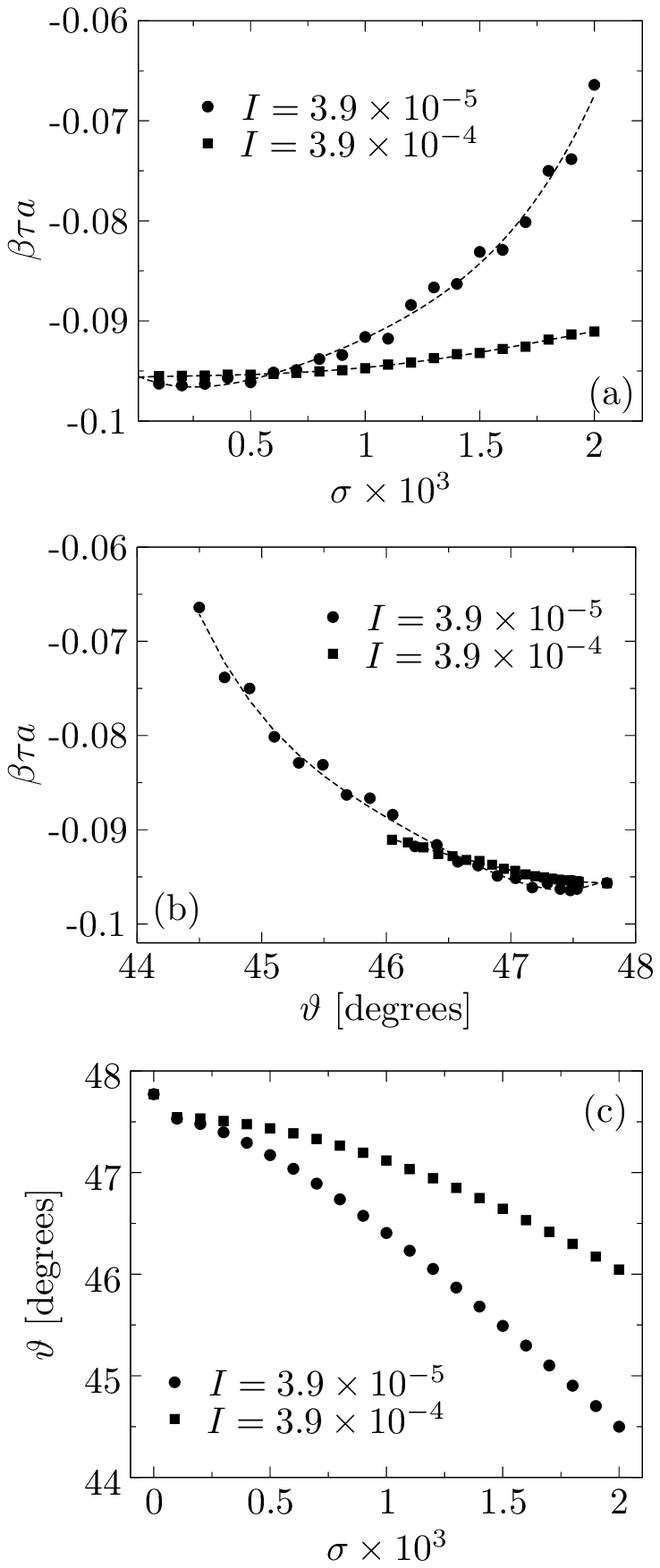}
\end{center}
\caption{Line tension $\tau$ as a function of the surface charge density
         $\sigma=\tilde\sigma a^2/e$ (a) and of the contact angle $\vartheta$
         (b) for $T^*=0.8T^*_c$ and $u_{\text{w}}/u=0.69$. 
         The two types of symbols correspond to distinct values of the ionic
         strength $I=\tilde Ia^3$ in the bulk liquid phase ($\bullet$ for
         $I=3.9\times10^{-5}$ ($\tilde I=1$mM) and $\scriptstyle\blacksquare$
         for $I=3.9\times10^{-4}$ ($\tilde I=10$mM)); the uncertainty
         of the line tension values corresponds approximately to the size of
         the symbols.
         $T^*=0.8\times T^*_c$ is below the wetting transitions, i.e., 
         $T^*<T_w^*(\sigma,I)$, for both ionic strengths $I$ and the whole
         range of surface charges $\sigma$ shown. 
         The dashed lines are guides to the eye. 
         The rightmost points in panel (a) correspond to the leftmost
         points in (b).
         See Appendix~\ref{ctau} for details concerning the calculation of
         the line tension.
         Panel (c) displays the contact angle $\vartheta$ as function of
         $\sigma$ with $I$ and $T^*$ fixed (route (ii)).}
\label{tauvssigma}
\end{figure}

Figure~\ref{tauvsT} shows the line tension for the case in which the contact 
angle is changed using route (i) for two distinct values of the ionic strength
$I$ and for a constant surface charge density $\sigma=1\times10^{-3}$ 
($\tilde\sigma=0.1\mu$C/cm$^2$). 
According to Fig.~\ref{tauvsT}(a), below the wetting transition the
line tension $\tau(T^*,\sigma,I)<0$ is a monotonically increasing function of the temperature 
$T^*<T_\text{w}^*(\sigma,I)$.
Consequently, since $T_\text{w}^*(\sigma,I)$, and thus the deviation
$T_\text{w}^*(\sigma,I)-T^*$ from the wetting temperature, increases upon 
increasing $I$ for fixed $\sigma$ (see Fig.~5 in Ref.~\cite{Ibagon}), the line
tension $\tau(T^*,\sigma,I)<0$ decreases upon increasing $I$ for fixed $T^*$ and
$\sigma$.
Moreover, as discussed in Fig.~\ref{tauvssigma}(a) below, the line 
tension $\tau(T^*,\sigma,I)<0$ is a monotonically increasing function of the
surface charge density $\sigma>0$ for fixed $T^*$ and $I$.
Therefore, the line tension $\tau(T^*,\sigma,I)$ of the pure, salt-free 
($I=0$) solvent in contact with a neutral ($\sigma=0$) wall, which is also
shown in Fig.~\ref{tauvsT}, can be larger or smaller than the one
for the cases $I>0,\sigma>0$.
  
The line tension is negative and its strength decreases upon decreasing the contact angle 
(see Fig.~\ref{tauvsT}(b)), which is in line with the predictions of the IDM
\cite{IndekeuLT1} for the case of first-order wetting transitions for 
charge-free fluids with short-ranged interactions.
The absolute value of the line tension is larger for the higher ionic
strength $I=3.9\times10^{-4}$ ($\tilde I=10$mM) at fixed temperature.  
We have not considered smaller contact angles because they require larger
system sizes and therefore generate substantially higher computational costs.
According to Ref.~\cite{IndekeuLT1}, the line tension in the case of 
first-order wetting transitions of fluids with short-ranged interactions are
expected to change sign from negative to positive upon decreasing the contact
angle $\vartheta$ and to be positive at the wetting transition temperature 
$T_\text{w}^*$, i.e., for $\vartheta=0$. 
This agrees also with the results reported in Refs.~\cite{Getta1998, Bauer} 
for long-ranged forces. 
Our data do not allow us to confirm this prediction, but one can infer from 
the available data that such a change in sign is rather plausible. 
In this case, the asymptotic behavior of $\tau$ for $\vartheta\to0$ predicted
in Ref.~\cite{IndekeuLT1} is given by 
$\tau\sim\tau_\text{w}+c_1\vartheta\ln\vartheta+c_2\vartheta+\mathcal{O}
(\vartheta^2)$.

\begin{figure}[!t]
\begin{center}
\includegraphics[scale=0.8]{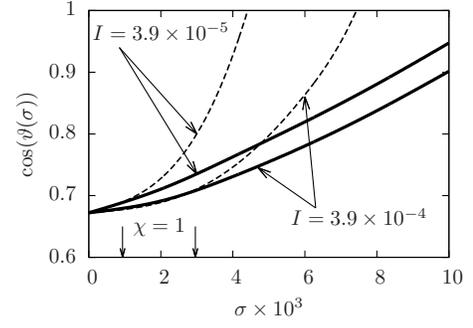}
\end{center}
\caption{Comparison of $\cos(\vartheta(\sigma))$ for the systems discussed
         in Fig.~\ref{tauvssigma} (solid lines) with the asymptotic expression
         given by Eq.~(\ref{eq:costhetaasym}) (dashed lines) derived in 
         Refs.~\cite{Kang2,Dorr2014} for temperature $T^*=0.8T_c^*$, surface
         charge density $0\leq\sigma\leq0.01$ ($0\leq\tilde\sigma\leq 
         1\,\mu\text{C}/\text{cm}^2$), and ionic strength 
         $I\in\{3.9\times10^{-5},3.9\times10^{-4}\}$ 
         ($\tilde I\in\{1\,\text{mM},10\,\text{mM}\}$).
         The asymptotic expressions apply up to surface charges $\sigma$
         for which the dimensionless parameter 
         $\chi=|\sigma|\sqrt{2\pi l_B/(\epsilon I)}$ attains unity, which 
         is marked by the vertical arrows.}
\label{costheta}
\end{figure}

Figure \ref{tauvssigma} shows the line tension for the case that the contact
angle is varied by using route (ii) for two values of the ionic strength 
$I$ and for $T^*=0.8T^*_c$. 
The line tension $\tau(T^*,\sigma,I)<0$ is a monotonically increasing function of the
surface charge density $\sigma>0$, and, as already discussed above in 
connection with Fig.~\ref{tauvsT}(a), it increases upon decreasing the
ionic strength $I$.
Here, only small surface charge values 
($\sigma=1\times10^{-4}-2\times10^{-3}$) have been considered. 
Accordingly, small contact angles, which correspond to large surface charges,
have not been studied. 
The technical reason for this is that in order to avoid contributions from the
corners of the integration box, these corners should be located far away from
all interfaces such that the density profiles near the corners attain their
bulk values (see Appendix \ref{ctau} and Sec. \ref{LTC}). 
Achieving this for small contact angles is more difficult in the case of the
electrolyte solution than for the pure solvent, mainly due to the density 
distributions of the ions. 
Figures~\ref{densitymaps1} and \ref{densitymaps2} show density distributions
of the solvent $\rho_0(\bar x, \bar z)$ and of the ions $\rho_\pm(\bar 
x, \bar z)$ for $\sigma=1\times10^{-4}$ ($\tilde\sigma=0.01\mu$C/cm$^2$)
and $\sigma=8\times10^{-3}$ ($\tilde \sigma=0.8\mu$C/cm$^2$), 
respectively. 
Both for Fig.~\ref{densitymaps1} and Fig.~\ref{densitymaps2}, the
bulk densities of the ions are $\rho_{\pm}=I=3.9\times10^{-5}$ 
($\tilde I=1$mM). For $\sigma=1\times10^{-4}$, in Fig.~\ref{densitymaps1} 
one can see that for the positive ions in the
liquid phase the density profile attains its bulk value only in a 
small portion of 
the calculation
box, which makes it difficult to use the integration box  shown 
in , c.f., Fig.~\ref{Figctau} 
and to carry out the procedure
described in Appendix \ref{ctau} for the calculation of the line tension. 
Moreover one can see in, c.f.,
Fig.~\ref{taufluctuation}, which shows examples of the dependence of the 
estimator $\mathcal{T}(\mathcal{B}^{(1)},\mathcal{B}^{(2)})$ of the line 
tension (see Appendix~\ref{ctau}) on the box size for two different
surface charge densities, that the amplitude of the variations of the value of
$\mathcal{T}(\mathcal{B}^{(1)},\mathcal{B}^{(2)})$, i.e., the uncertainty
of the value of the line tension, increases if the surface charge 
density
$\sigma$ increases. Figure~\ref{tauvssigma}(a) shows that for small surface 
charge densities ($\sigma\lesssim1\times10^{-3}$) the
value of the line tension, the uncertainty of which corresponds
approximately to the size of the symbols, is, within the precision of the method
described in Appendix~\ref{ctau}, independent of the ionic strength $I$. 
However, as the surface charge density increases, the absolute value of the 
line tension $\tau$ decreases stronger for $I=3.9\times10^{-5}$ ($\tilde
I=1$mM) than for $I=3.9\times10^{-4}$ ($\tilde I=10$mM). 
This is related to the fact that due to screening for $I=3.9\times10^{-4}$ 
($\tilde I=10$mM) a larger  surface charge is needed to produce the same 
contact angle as for $I=3.9\times10^{-5}$ ($\tilde I=1$mM) (see 
Fig.~\ref{tauvssigma}(c)).
Thus upon increasing $I$, according to route (iii), the contact angle 
$\vartheta$ increases and so does the strength of the line tension.

Within the approximation of a field-free gas phase the asymptotic behavior
\begin{align}
   \cos(\vartheta(\sigma)) \simeq \cos(\vartheta(0)) +
   \frac{\sigma^2}{\gamma_{l,g}}\sqrt{\frac{\pi l_B}{2\epsilon I}}
   \left(\frac{\pi}{\vartheta(\sigma)}-1\right)
   \label{eq:costhetaasym}
\end{align}
has been derived in Refs.~\cite{Kang2,Dorr2014} for the case of the dimensionless
quantity $\dps\chi := |\sigma|\sqrt{\frac{2\pi l_B}{\epsilon I}}$ being small
($\chi\ll1$).
Figure~\ref{costheta} compares the curves $\cos(\vartheta(\sigma))$ for the
systems discussed in Fig.~\ref{tauvssigma} (solid lines) with the corresponding
asymptotic form Eq.~(\ref{eq:costhetaasym}) (dashed lines).
The asymptotic expressions are reliable up to surface charge densities $\sigma$
for which $\chi\approx1$, marked by the vertical arrows in Fig.~\ref{costheta}.
Hence Eq.~(\ref{eq:costhetaasym}) applies to small surface charge densities not
only in the case of the electric field being confined to a wedge-shaped liquid
phase, as in Refs.~\cite{Kang2,Dorr2014}, but also in the case of a non-vanishing
electric field in the gas phase.

\subsection{Density distributions close to the three-phase contact line}

\begin{figure*}[!t]
\begin{center}
\includegraphics[scale=0.95]{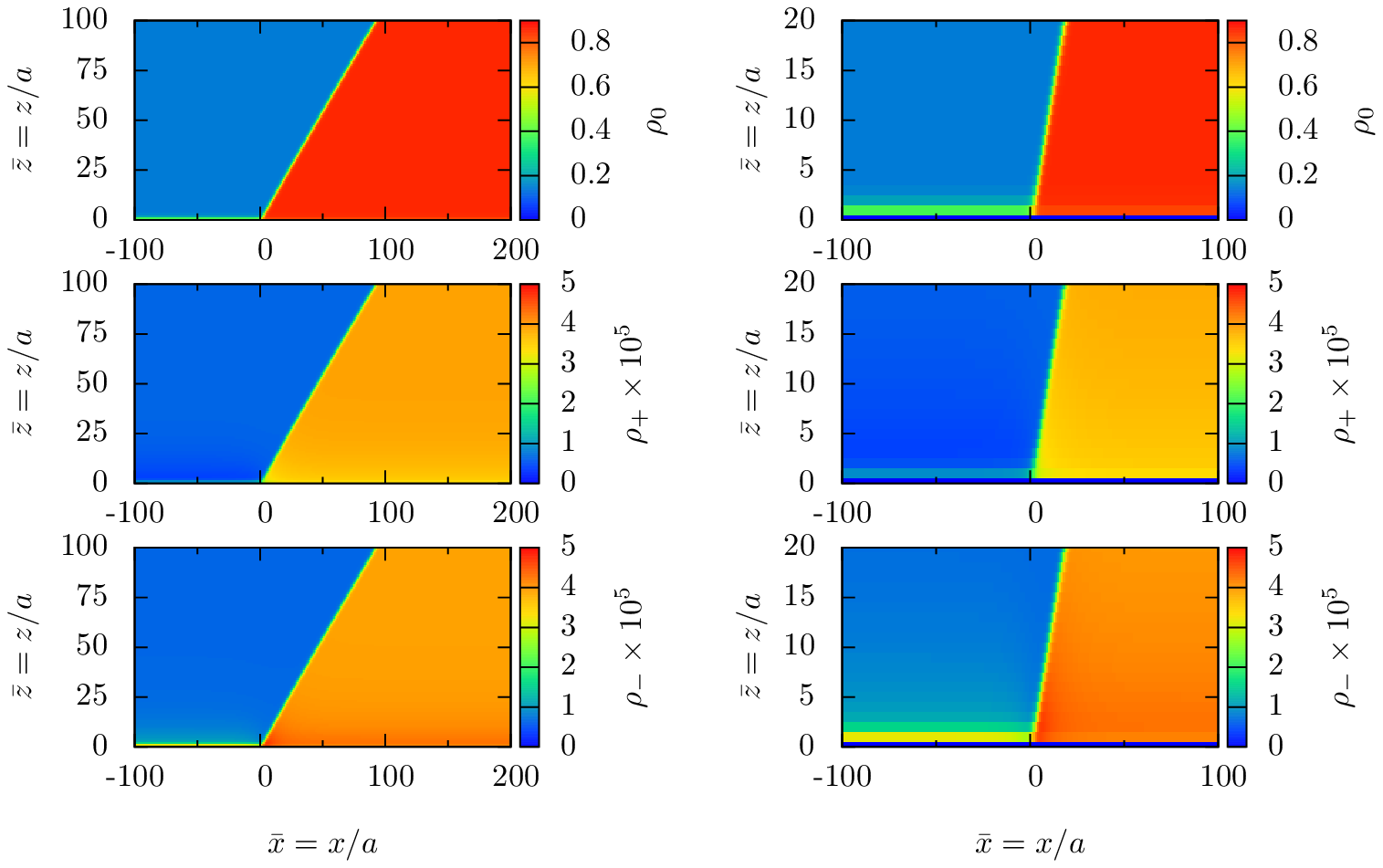} 
\end{center}
\caption{Density distributions of the solvent $\left[\rho_0(\bar x, 
         \bar z)\right]$, the cations $\left[\rho_+(\bar x, \bar z)
         \right]$, and the anions $\left[\rho_-(\bar x, \bar z)\right]$
         for $\sigma=1\times10^{-4}$ ($\tilde\sigma=0.01\mu$C/cm$^2$). 
         The bulk values of the density distribution of the cations and
         the anions are $\rho_\pm = I = 3.9\times10^{-5}$ ($\tilde I = 1$mM);
         $\rho_g=0.14$ and $\rho_l=0.86$ are the bulk values for the gas and
         the liquid, respectively.
         The contact angle is $\vartheta\approx47.5^\circ$. 
         The substrate is positively charged. 
         Therefore there is a high (low) density of negative (positive)
         ions in its vicinity. 
         The panels on the right show close-ups of the plots on the left in
         the vicinity of the TPCL.}
\label{densitymaps1}
\end{figure*}

\begin{figure*}[!t]
\begin{center}
\includegraphics[scale=0.95]{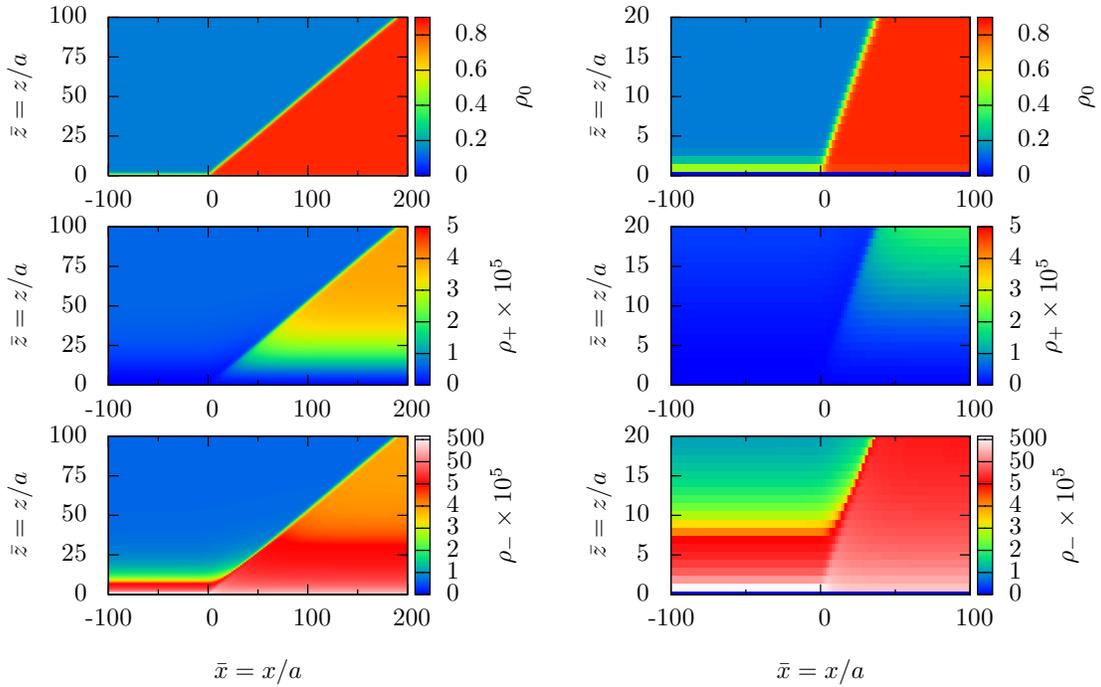}
\end{center}
\caption{Same as Fig.~\ref{densitymaps1} for $\sigma=8\times10^{-3}$ 
         ($\tilde\sigma=0.8\mu$C/cm$^2$). 
         The contact angle is $\vartheta\approx28.3^\circ$. 
         Note that in this case the density distribution $\rho_+$
         of the cations in the liquid phase needs more space in order to  
         attain its bulk value $I=3.9\times10^{-5} $($\tilde I = 1$mM) 
         than for  $\sigma=1\times10^{-4}$ ($\tilde \sigma=0.01\mu$C/cm$^2$)
         (see Fig.~\ref{densitymaps1}).}
\label{densitymaps2}
\end{figure*}

\begin{figure*}[!t]
\begin{center}
\includegraphics[scale=0.95]{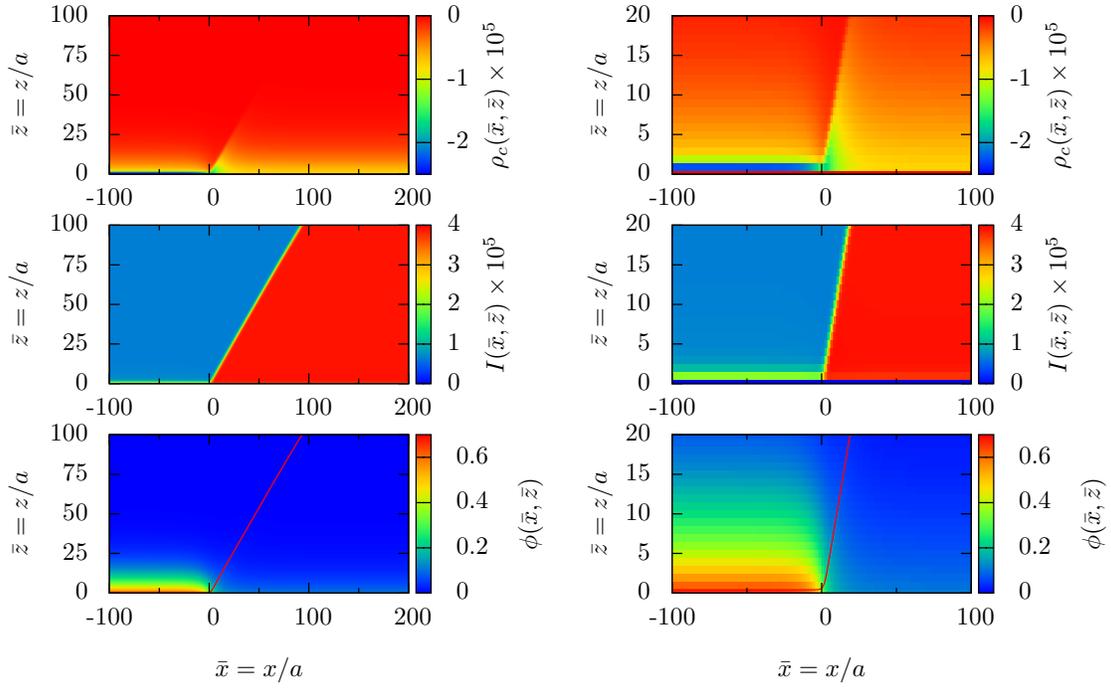}
\end{center}
\caption{Charge density $\rho_c(\bar x, \bar z)=\rho_+(\bar x, \bar z)-
         \rho_-(\bar x, \bar z)$, local ionic strength $I(\bar x, \bar z)=
         \tfrac{1}{2}\left(\rho_+(\bar x,\bar z)+\rho_-(\bar x, \bar 
         z)\right)$,  and electrostatic potential $\phi(\bar x, \bar z)=\beta
         e\tilde\phi(\bar x, \bar z)$ for $\sigma=1\times10^{-4}$ 
         ($\tilde\sigma=0.01\mu$C/cm$^2$). 
         The contact angle is $\vartheta\approx47.5^\circ$. 
         The red line in the bottom panels for the electrostatic 
         potential indicates the shape $\ell(\bar x)$ of the liquid-gas 
         interface obtained from Eq.~(\ref{lx}). 
         The panels on the right show close-ups of the plots on the left 
         in the vicinity of the TPCL.}
\label{s1e-4}
\end{figure*}

\begin{figure*}[!t]
\begin{center}
\includegraphics[scale=0.95]{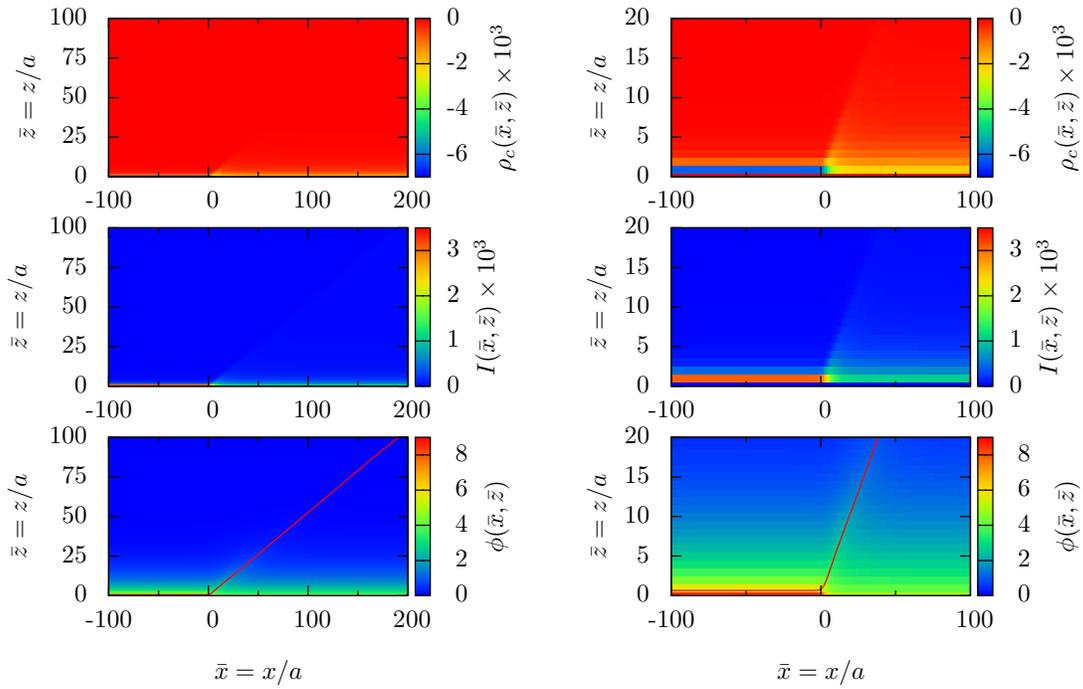}
\end{center}
\caption{Same as Fig.~\ref{s1e-4} for $\sigma=8\times10^{-3}$ 
         ($\tilde\sigma=0.8\mu$C/cm$^2$). 
         The contact angle is $\vartheta\approx28.3^\circ$.}
\label{s8e-3}
\end{figure*}

The microscopic structure of the electrolyte solution close to the TPCL is 
illustrated via density
maps in Figs.~\ref{densitymaps1} and \ref{densitymaps2} for 
$I=3.9\times10^{-5}$ ($\tilde I=1$mM), 
$T^*=0.8T^*_c$, and two
values of the surface charge density: $\sigma=1\times10^{-4}$($\tilde 
\sigma=0.01\mu$C/cm$^2$) (see  Fig.~\ref{densitymaps1})
and $\sigma=8\times10^{-3}$ ($\tilde \sigma=0.8\mu$C/cm$^2$) (see  
Fig.~\ref{densitymaps2}) . The contact angles 
are
$\vartheta\approx47.5^\circ$ for $\sigma=1\times10^{-4}$ and 
$\vartheta\approx28.3^\circ$ for
$\sigma=8\times10^{-3}$. Apart from the difference in contact angle and
from the different densities
of anions and cations in the vicinity of the wall due to the difference in 
surface charge density
$\sigma$, one can infer that for larger values of the surface charge   
the density distributions of the ions differ significantly from their
bulk values over larger distances from the substrate. 
The anion densities $\rho_-(z)$ close to the gas-wall interface in
Fig.~\ref{densitymaps2} are large because in the present study 
(see Sec.~\ref{model}) we assume a laterally uniform surface
charge density, which is not modified by charge regulation.
For setups with surface charge densities being determined by charge 
regulation, significantly smaller surface charge densities would
occur and hence smaller ion densities $\rho_\pm(z)$ close to the gas-wall
interface.

Figures \ref{s1e-4} and \ref{s8e-3} show the charge
density $\rho_c(\bar x, \bar z)=\rho_+((\bar x,
\bar z)-\rho_-(\bar x,\bar z)$, the local ionic strength $I(\bar x, \bar
z)=\tfrac{1}{2}\left(\rho_+((\bar x,
\bar z)+\rho_-((\bar x,\bar z))\right)$, and the electrostatic potential 
$\phi(\bar x, \bar z)=\beta
e\tilde\phi(\bar x,\bar z)$ for
the same set of parameters as in Figs.~\ref{densitymaps1} and
\ref{densitymaps2}, respectively. For a 
small surface charge density $\sigma$ (see
Fig.~\ref{s1e-4}) the charge density $\rho_c(\bar x,\bar z)$ has a region in the gas close to the
liquid-gas interface where $\rho_c(\bar x,
\bar z)$ is less negative than $\rho_c(-\infty,\bar z)$. If one takes a path 
parallel to the surface at small $\bar z$ from the gas side, the
charge
density $\rho_c(\bar x,
\bar z)$ is quasi constant in the gas phase far away from the liquid-gas 
interface, increases upon
approaching the liquid-gas interface from the gas side, drops to a rather 
low value on the liquid
side of the liquid-gas interface, and ultimately increases
towards a constant value in the liquid phase. This charge separation in the vicinity of the
liquid-gas interface and of the TPCL is caused by the variation of
the local permittivity of the
solvent $\varepsilon(\rho_0(\bar x, \bar z))$ which is higher in the liquid phase (see Eq.
\ref{epsi2D}). On the other hand, the structure of the local ionic strength 
distribution $I(\bar x, \bar z)$, which for constant $\bar z$ 
interpolates from the value in
the
gas phase to the value in the liquid phase, is almost independent of $\bar z$ 
within each phase. The electrostatic
potential $\phi(\bar x,\bar z)$, which is related to the charge density through 
Poisson's equation (Eq.~(\ref{PE2D})), does not follow the liquid-gas 
interface in that the equipotential lines bend away from it. Moreover, 
there is an
electrostatic potential difference between the liquid and the gas phase in the 
vicinity of the
TPCL.
For a large surface charge density $\sigma$  (see Fig.~\ref{s8e-3}), the high
charge density $\rho_c$ in the vicinity of 
the substrate on the gas side screens the surface 
charge of
the substrate within a few layers. This is in contrast to the case of
small surface charge for which the charge density
$\rho_c(\bar x, \bar z)$ approaches its vanishing bulk value more slowly
(compare Fig.~\ref{s1e-4}). This different behavior is due to the
nonlinear character of Poisson's equation (see Eq.~(\ref{PE2D})); for
small values of the
surface charge density $\sigma$ its solution is close to
the
solution of the linearized equation in which the number densities of the ions
decay exponentially to their bulk values on the scale of the Debye
 length 
$\kappa$ of the bulk phase. In contrast, for
large surface charge density $\sigma$  both the density distributions of 
the ions and the
electrostatic potential $\phi$ deviate significantly from the linear solution in the vicinity of the
substrate, and the exponential decay is only valid far away from it. For 
this large value of $\sigma$, the aforementioned nonmonotonic variation 
of $\rho_c(\bar x, \bar z)$ in the vicinity of the liquid-gas interface from
the gas side is not observed. However, $\rho_c(\bar x,\bar z)$ 
becomes more negative in the vicinity of the liquid-gas
interface
from the liquid side. This qualitative difference as a function of 
$\sigma$ in the behavior of the charge density in the vicinity of the
TPCL  results in a different behavior of the electrostatic potential. For 
large $\sigma$ the difference of the values of the
electrostatic potential in the liquid and in the gas phase is not as pronounced
as for smaller surface
charge densities (see Fig.~\ref{s1e-4}). For all $\sigma$, far away from 
the substrate the charge density $\rho_c(\bar x, \bar z)$ and the electrostatic
potential $\phi(\bar x,\bar z)$ vanish and the local ionic strength 
attains its bulk value, here $I=3.9\times10^{-5}$. 

\clearpage

\section{Conclusions and summary}\label{CS}
We have investigated the line tension and the structure of the
three-phase contact line (Fig.~\ref{lxsketch}) of an electrolyte
solution in contact with a charged substrate by using 
density functional theory applied to a lattice model \cite{Ibagon}. For the
pure, i.e., salt-free solvent, the equilibrium shape of the liquid-gas 
interface approaches its
asymptotes from above, as expected for systems exhibiting second-order 
wetting transitions (Fig.~\ref{CPS}). Near the wetting transition the 
line tension vanishes proportional to the contact 
angle (Fig.~\ref{ltps}) which itself goes to zero at the wetting 
transition temperature.
For the electrolyte solution, the equilibrium shape of the liquid-gas 
interface approaches its
asymptote from below as expected for systems exhibiting first-order 
wetting transitions (Fig.~\ref{ES}). 
If the contact angle is changed by varying the temperature while 
keeping the surface charge fixed, the line tension becomes less negative 
as the temperature is increased (Fig.~\ref{tauvsT}(a)), i.e.,
as the
contact angle is decreased. For fixed temperature, the line tension is  
more negative for the larger ionic
strength (Fig.~\ref{tauvsT}(a)). If the contact angle is changed by 
varying the surface charge
density at fixed temperature, the line
tension becomes less negative as the surface charge is increased 
(Fig.~\ref{tauvssigma}(a)). For small surface charges this 
decrease of the strength of the
line tension depends only weakly on
the ionic strength (Fig.~\ref{tauvssigma}(a)). However, for larger surface 
charges the
decrease of the strength of the line tension is steeper for the smaller ionic 
strength (Fig.~\ref{tauvssigma}(a)). We have also calculated the intrinsic
equilibrium structure of the three-phase contact line for various charge 
densities. For
large surface charge densities, nonlinear effects of the
Poisson-Boltzmann theory dominate. This results in distributions of 
the ions and of the electrostatic
potential which differ from those for small surface charge densities 
(Figs.~\ref{densitymaps1}, \ref{densitymaps2}, \ref{s1e-4}, and \ref{s8e-3}).

On the one hand, technically the lattice model facilitates the
reliable determination of these structures and properties. 
On the other hand, using a lattice model causes a difficulty
for calculating the line 
tension, because within this model the 
liquid-gas surface tension depends on the orientation of the
interfacial plane relative to the
underlying lattice. Accordingly, this aspect of our study 
should be regarded as a
first step towards the microscopic calculation of line tensions in 
electrolyte solutions and should be
compared with not yet available results from continuum models for 
electrolytes. 
Moreover, for technical reasons the asymptotic
behavior of the line tension upon
approaching the wetting
transition and the influence of a large surface charge densities of the 
substrate on the line tension
could
not be addressed within the present approach; they deserve to be 
analyzed in the future within continuum models.

\vfill 

\appendix

\section{Line tension calculation within the lattice model} \label{ctau}

\begin{figure*}[!t]
\begin{center}
\includegraphics{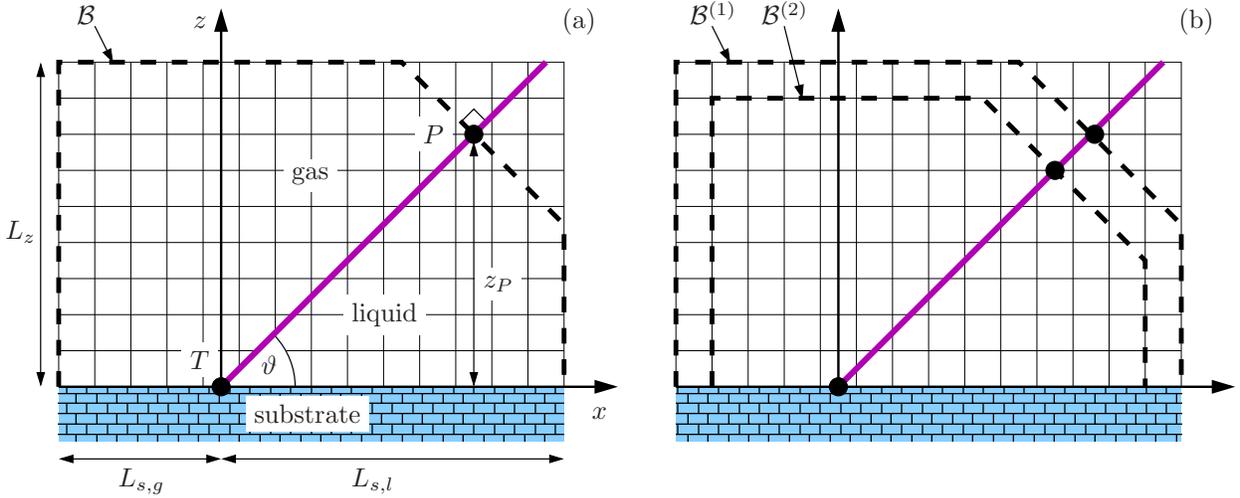}
\end{center}
   \caption{(a) Underlying geometry for calculations of the line
            tension. 
            The Euler-Lagrange equations in Eq.~(\ref{ELE2D}) are solved via 
            an iterative algorithm in a rectangular box which includes
            as a subset the computational box $\mathcal{B}$ (dashed lines)
            which is used to determine the line tension.
            $\mathcal{B}$ encloses the three-phase contact line $T$ and its
            boundaries cut perpendicularly through all interfaces.
            It is characterized by the lengths $L_{s,g}$ and $L_{s,l}$
            of the substrate-gas and the substrate-liquid interface at
            the plane $z=0$, respectively, by the width $L_z$ in 
            $z$-direction, and by the $z$-coordinate $z_P$ of the point $P$, 
            where $\mathcal{B}$ intersects the liquid-gas interface (thick 
            inclined magenta line, forming the contact angle 
            $\vartheta$ with the substrate surface).
            The convention of the substrate-fluid interfaces being located
            at $z=0$ affects not only the geometrical quantities $L_{s,g}$,
            $L_{s,l}$, and $A(\mathcal{B})$, but also the definition of the
            substrate-gas and the substrate-liquid surface tensions 
            $\gamma_{s,g}$ and $\gamma_{s,l}$, respectively.
            The system is translationally invariant in the $y$-direction. 
            Note that the boundary cutting through the liquid-gas interface
            crosses some of the cells, leaving only a fraction of
            each of them inside $\mathcal{B}$; it is this fraction with
            which these cells contribute to the total grand canonical free
            energy.
            (b) The line tension is determined via the estimator
            $\overline{\mathcal{T}}(\mathcal{B}^{(1)}, \mathcal{B}^{(2)})$ in
            Eq.~(\ref{eq:barT}), which takes two different calculational boxes
            $\mathcal{B}^{(1)}$ and $\mathcal{B}^{(2)}$ as its arguments 
            (see also Fig.~\ref{taufluctuation}).}
\label{Figctau}
\end{figure*}

For the line tension calculation, computational boxes $\mathcal{B}$
have been used which cut perpendicularly through all interfaces and
which are bounded by the substrate-fluid interfaces being located at $z=0$
(Fig.~\ref{Figctau}(a)). 
As discussed in Ref.~\cite{SND}, this type of boxes ensures that, for
sufficiently large $\mathcal{B}$ and within continuum models, no 
artificial contributions to the grand canonical free energy 
$\Omega(\mathcal{B})$ appear, which are due to the edges of 
$\mathcal{B}$ or due to inhomogeneities caused by the boundaries of 
$\mathcal{B}$.
According to Eq.~(\ref{lteq}), the grand canonical free energy 
$\Omega(\mathcal{B})$ of $\mathcal{B}$ per length $L$ of the straight
three-phase contact line $T$ (see Fig.~\ref{Figctau}) is given by
\begin{align}
   \frac{\Omega(\mathcal{B})}{L} =\ 
   &\Omega_b A(\mathcal{B}) + \gamma_{l,g}(\mathcal{B})L_{l,g}(\mathcal{B}) 
   + \gamma_{s,l} L_{s,l}(\mathcal{B}) \notag\\
   &+ \gamma_{s,g}L_{s,g}(\mathcal{B}) + \tau,
   \label{eq:OmB}
\end{align}
where $\Omega_b=\Omega_g=\Omega_l=-p$ is the density of the bulk grand 
potential, i.e., the negative pressure, given by Eq.~(\ref{bulk1}) evaluated at
the equilibrium densities, $A(\mathcal{B})$ is the cross-sectional area of
$\mathcal{B}$, such that $V(\mathcal{B})=A(\mathcal{B})L$ is the volume of the
fluid inside $\mathcal{B}$, $L_{l,g}(\mathcal{B}) =
z_P(\mathcal{B})/\sin(\vartheta)$ is the length of the intersection of the 
liquid-gas interface inside $\mathcal{B}$ with the $x$-$z$-plane (see the 
thick magenta line $TP$ in Fig.~\ref{Figctau}), and $L_{s,l}(\mathcal{B})$ and 
$L_{s,g}(\mathcal{B})$ are the linear extensions of the substrate-liquid and 
the substrate-gas interface in the $x$-direction, respectively.
The substrate-liquid surface tension $\gamma_{s,l}$ and the 
substrate-gas surface tension $\gamma_{s,g}$ in Eq.~(\ref{eq:OmB}) do
not depend on $\mathcal{B}$ and they can be inferred from the substrate 
being in contact with the bulk liquid and bulk gas, respectively.
Note that the quantities $A(\mathcal{B})$, $\gamma_{s,l}$, 
$L_{s,l}(\mathcal{B})$, $\gamma_{s,g}$, and $L_{s,g}(\mathcal{B})$ depend on
the choice of the convention for the substrate-fluid interface 
position (here $z=0$, see Fig.~\ref{Figctau}(a)), so that the line tension 
$\tau$ in Eq.~(\ref{eq:OmB}) depends on this choice of the convention, 
too.

A difference between continuum and lattice models arises
with respect to $\gamma_{l,g}(\mathcal{B})$ in Eq.~(\ref{eq:OmB}): Within 
continuum models, $\gamma_{l,g}(\mathcal{B})=\gamma_{l,g}^{(0)}$ is 
independent of $\mathcal{B}$ and it coincides with the liquid-gas interfacial 
tension $\gamma_{l,g}^{(0)}$, whereas within lattice models 
$\gamma_{l,g}(\mathcal{B})$ varies with $\mathcal{B}$ since the tilted free 
liquid-gas free interface  (see the thick magenta line $TP$ in 
Fig.~\ref{Figctau}), which is inclined by the contact angle $\vartheta$ with 
respect to the substrate, in general does not match the underlying lattice 
grid. 

In order to estimate the line tension $\tau$ in Eq.~(\ref{eq:OmB}), the 
contribution $\gamma_{l,g}(\mathcal{B})L_{l,g}(\mathcal{B})$ in 
Eq.~(\ref{eq:OmB}) is written in the form 
\begin{align}
   \gamma_{l,g}(\mathcal{B})L_{l,g}(\mathcal{B}) 
   = \frac{\gamma_{l,g}^{(0)}z_P(\mathcal{B})}{\sin(\vartheta)} 
     + \delta_{l,g}(\mathcal{B})
   \label{eq:fluc}
\end{align}
with $\vartheta$ independent of $\mathcal{B}$.
Being maximally ignorant of the relative position of the liquid-gas 
interface with respect to the lattice grid, the probabilities of finding 
positive or negative deviations $\delta_{l,g}(\mathcal{B})$ are equal such 
that the expectation value $\langle\delta_{l,g}(\mathcal{B})\rangle$ vanishes.
Consequently, according to Eq.~(\ref{eq:OmB}), the quantity
\begin{align}
   \mathcal{T}(\mathcal{B}) :=\ 
   &\frac{\Omega(\mathcal{B})}{L} - \Omega_b A(\mathcal{B}) 
   - \frac{\gamma_{l,g}^{(0)}z_P(\mathcal{B})}{\sin(\vartheta)}
   \notag\\
   &- \gamma_{s,l} L_{s,l}(\mathcal{B}) - \gamma_{s,g}L_{s,g}(\mathcal{B}) 
   \label{eq:Tcaldef}
\end{align}
is expected to vary, within the set of computational boxes $\mathcal{B}$
of the type specified above, as a function of $\mathcal{B}$ around the line 
tension $\tau$ according to
\begin{align}
   \mathcal{T}(\mathcal{B}) = \tau + \delta_{l,g}(\mathcal{B}).
   \label{eq:Tcaltau}   
\end{align}
In principle, Eq.~(\ref{eq:Tcaltau}) facilitates to determine
the line tension $\tau$ as the $\mathcal{B}$-independent ``background''
contribution to $\mathcal{T}(\mathcal{B})$.
However, $\mathcal{T}(\mathcal{B})$ depends sensitively on the value 
$\vartheta$ of the contact angle, which turns out to be difficult to 
track with the necessary numerical precision.
A possible approach to determine the line tension $\tau$ without precise
knowledge of the contact angle $\vartheta$ consists of the following:
Consider two computational boxes $\mathcal{B}^{(1)}$ and $\mathcal{B}^{(2)}$
with $z_P(\mathcal{B}^{(1)})=:z_1$ and $z_P(\mathcal{B}^{(2)})=:z_2$.
The contributions $\sim 1/\sin(\vartheta)$ from Eq.~(\ref{eq:Tcaldef}) cancel
in the combination $z_1\mathcal{T}(\mathcal{B}^{(2)})-
z_2\mathcal{T}(\mathcal{B}^{(1)})$ so that instead of Eq.~(\ref{eq:Tcaltau})
one can use the expression
\begin{align}
   \overline{\mathcal{T}}(\mathcal{B}^{(1)},\mathcal{B}^{(2)}) 
   :=&\
   \frac{z_1\mathcal{T}(\mathcal{B}^{(2)})-z_2\mathcal{T}(\mathcal{B}^{(1)})}
        {z_1-z_2}
   \notag\\
   =&\ \tau + 
   \frac{z_1\delta_{l,g}(\mathcal{B}^{(2)})-z_2\delta_{l,g}(\mathcal{B}^{(1)})}
        {z_1-z_2}
   \label{eq:barT}
\end{align}
in order to infer the line tension $\tau$ as that contribution to 
$\overline{\mathcal{T}}(\mathcal{B}^{(1)},\mathcal{B}^{(2)})$, which is 
independent of $\mathcal{B}^{(1)}$ and $\mathcal{B}^{(2)}$.

\begin{figure*}[!t]
\begin{center}
\includegraphics[scale=1,clip=true]{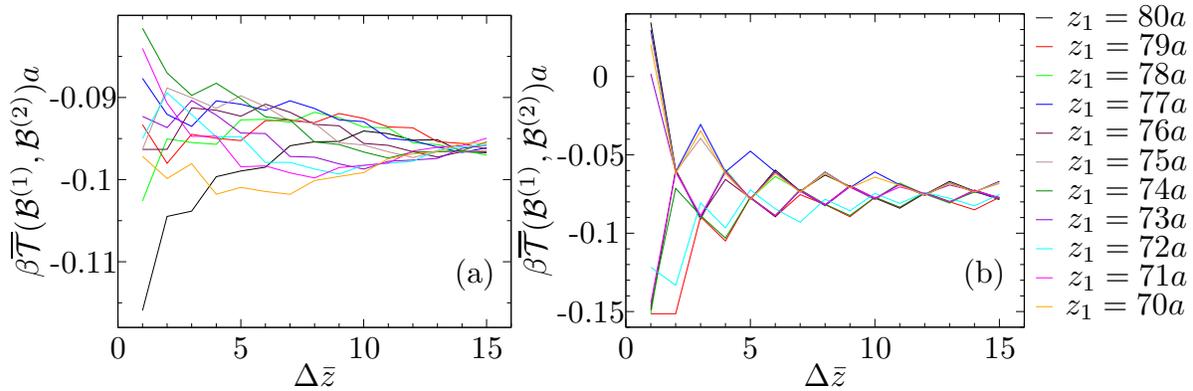}
\end{center}
   \caption{Data for $\overline{\mathcal{T}}(\mathcal{B}^{(1)},
            \mathcal{B}^{(2)})$ introduced in Eq.~(\ref{eq:barT}) for
            an electrolyte solution with $I=3.9\times10^{-5}$ ($\tilde I= 1mM$)
            at $T^*=0.8\times T^*_c$ for $\sigma=5\times10^{-4}$ ($\tilde
            \sigma=0.05\mu$C/cm$^2$) (a) and $\sigma=1.8\times10^{-3}$ 
            ($\tilde \sigma=0.18\mu$C/cm$^2$) (b). 
            The various colors correspond to different computational boxes
            $\mathcal{B}^{(1)}$ in Eq.~(\ref{eq:barT}), i.e., they
            correspond to different distances $z_1 =
            z_P(\mathcal{B}^{(1)})$ of the point $P$, where the boundary
            of the box intersects the liquid-gas interface, from the 
            substrate (see Fig.~\ref{Figctau}(a)). 
            The expression $\overline{\mathcal{T}}(\mathcal{B}^{(1)},
            \mathcal{B}^{(2)})$ is shown as a function of $\Delta \bar z
            = \bar z_1-\bar z_2$ with $z_2=z_P(\mathcal{B}^{(2)})$ (see
            Fig.~\ref{Figctau}(b)).
            The line tension $\tau$ is inferred from these plots as that
            contribution to $\overline{\mathcal{T}}(\mathcal{B}^{(1)},
            \mathcal{B}^{(2)})$, which is constant, i.e., independent of
            $z_1$ and $\Delta z$ (see Eq.~(\ref{eq:barT})).}
\label{taufluctuation}
\end{figure*}

We have calculated the expression 
$\overline{\mathcal{T}}(\mathcal{B}^{(1)},\mathcal{B}^{(2)})$ in 
Eq.~(\ref{eq:barT}) by fixing the intersection of box $\mathcal{B}^{(1)}$ with
the liquid-gas interface at wall distances $z_1\in[30a,40a]$ for the pure 
solvent and at wall distances $z_1\in[70a,80a]$ for the electrolyte
solution.
The size of box $\mathcal{B}^{(2)}$ has been varied accordingly
such that $\Delta z:=z_1-z_2\in\{a,2a,\cdots,15a\}$. 
This procedure has been repeated for all integers $z_1$ in the 
corresponding intervals for the pure solvent and for the electrolyte 
solution. 
The values of $L_{s,l}$ and $L_{s,g}$ (see Fig.~\ref{Figctau}) are
determined via the position at which the asymptote of the gas-liquid
interface intersects the plane $z=0$. 
The size of the rectangular box used to determine the equilibrium 
profiles depends on the contact angle $\vartheta$, i.e., for smaller contact
angles a larger extension in the $x$-direction is needed. 
For the pure solvent, as the smaller size we have used 
$L_x\times L_y=300a\times60a$, while as the bigger size $1500a\times60a$ 
has been used. 
For the electrolyte solution a fixed box size of $400a\times100a$ was 
used. 
Figure~\ref{taufluctuation} shows the values of
$\overline{\mathcal{T}}(\mathcal{B}^{(1)},\mathcal{B}^{(2)})$ calculated
for an electrolyte solution with $I=3.9\times10^{-5}$ ($\tilde I= 1mM$), 
$T^*=0.8T^*_c,u_w/u=0.69$, and for two values of the surface 
charge density: $\sigma=5\times10^{-4}$ ($\tilde \sigma=0.05\mu$C/cm$^2$) 
(Fig.~\ref{taufluctuation}(a))  and $\sigma=1.8\times10^{-3}$ ($\tilde
\sigma=0.18\mu$C/cm$^2$) (Fig.~\ref{taufluctuation}(b)). 
The variation of $\overline{\mathcal{T}}(\mathcal{B}^{(1)},
\mathcal{B}^{(2)})$ with the size of both integration boxes, 
$\mathcal{B}^{(1)}$ and $\mathcal{B}^{(2)}$, is clearly visible.
Nonetheless $\overline{\mathcal{T}}(\mathcal{B}^{(1)}, \mathcal{B}^{(2)})$
is distributed around a specific value $\tau$. 
In order to determine this value $\tau$, which here is called the line
tension, that value of $\Delta z$ is chosen which 
renders the smallest variation in 
$\overline{\mathcal{T}}(\mathcal{B}^{(1)}, \mathcal{B}^{(2)})$ for 
different $z_1$; $\tau$ is taken to be the mean value of the smallest and
the largest values of $\overline{\mathcal{T}}(\mathcal{B}^{(1)}, 
\mathcal{B}^{(2)})$ for that particular choice $\Delta z$. 
We note that the amplitude of the variations in 
$\overline{\mathcal{T}}(\mathcal{B}^{(1)}, \mathcal{B}^{(2)})$ increase
with increasing $\sigma$, i.e., with decreasing contact angle $\vartheta$.
This can be inferred from the different scales on axes of ordinates 
in Figs.~\ref{taufluctuation}(a) and (b). 
The corresponding behavior is similar for the other values of the surface
charge density considered here. 
If the surface charge density is fixed and the contact angle 
$\vartheta$ is varied by changing the temperature, the amplitudes of the 
variations in $\overline{\mathcal{T}}(\mathcal{B}^{(1)}, 
\mathcal{B}^{(2)})$ increase upon increasing the temperature 
$T^*$, i.e., upon decreasing the contact angle $\vartheta$.

\end{document}